\documentclass{ws-ijgmmp}
\usepackage[utf8]{inputenc}
\usepackage[english]{babel}
\usepackage{graphicx}
\usepackage{bm}
\usepackage{verbatim}
\usepackage{hyperref}
\usepackage{mathrsfs}
\usepackage{upgreek}
\usepackage{physics}
\usepackage{amssymb}
\usepackage{amsmath}
\usepackage{subfig}
\usepackage{float}
\usepackage{booktabs}
\usepackage{bigints}

\begin{document}

\markboth{G. Barca, E. Giovannetti, G. Montani}
{Comparison of the Bianchi I Cosmology in the Polymer and GUP paradigms}

%
\catchline{}{}{}{}{}
%

\title{Comparison of the Semiclassical and Quantum Dynamics of the Bianchi I Cosmology in the Polymer and GUP Extended Paradigms}

\author{Gabriele Barca}
\address{Department of Physics, La Sapienza University of Rome, P.le Aldo Moro 5, 00185 Roma, Italy\\
corresponding author: \email{gabriele.barca@uniroma1.it}}
\author{Eleonora Giovannetti}
\address{Department of Physics, La Sapienza University of Rome, P.le Aldo Moro 5, 00185 Roma, Italy\\
\email{eleonora.giovannetti@uniroma1.it}}
\author{Giovanni Montani}
\address{Department of Physics, La Sapienza University of Rome, P.le Aldo Moro 5, 00185 Roma, Italy; Fusion and Nuclear Safety Department, European Nuclear Energy Agency (ENEA), C.R. Frascati, Via E. Fermi 45, 00044 Frascati, Italy\\
\email{giovanni.montani@enea.it}}

\maketitle

\begin{history}
\received{28 December 2021}
\accepted{15 February 2022}
{Published 12 April 2022}
\end{history}

\begin{abstract}
We compare in some detail Polymer Quantum Mechanics and the Generalized Uncertainty Principle approach to clarify to what extent we can treat them on the same footing. We show that, while on a semiclassical level they may be formulated as similar modifications of the Poisson algebra, on a quantum level they intrinsically differ because PQM implies no absolute minimal uncertainty on position.

\noindent Then we implement these schemes to Bianchi I cosmology on a semiclassical level deforming only the algebra of the Universe volume, searching for alternative formulations able to account for the modified Friedmann equations emerging in Brane Cosmology and Loop Quantum Cosmology. 

\noindent On a pure quantum level, we implement the two approaches through their original setups and reduce the two resulting Wheeler-DeWitt equations to the same morphological structure, showing how the polymer formalism is associated with a bouncing dynamics while in the Generalized Uncertainty Principle case the singularity is still present. The implications of the wavepacket spreading are also discussed in both approaches, outlining that, when the singularity survives, the Planckian era must necessarily be approached by a fully quantum (non-peaked) state of the Universe.
\end{abstract}


\section{Introduction}
The two most developed proposals to move classical General Relativity towards a quantum formulation are to be considered Loop Quantum Gravity (LQG) \cite{LQG} (see also covariant Loop Quantum Gravity \cite{cLQG}) and String Theories (ST) \cite{ST} (see \cite{STrev} for a recent review).

The cosmological implementation of LQG, known as Loop Quantum Cosmology (LQC) \cite{LQC,LQCrev}, has a quantum and semiclassical phenomenology well reproduced by the minisuperspace implementation of the so-called Polymer Quantum Mechanics (PQM) \cite{CorichiPQM}. Almost on the same level, the low energy phenomenology of ST can be interpreted as a modification of the Heisenberg Uncertainty Principle, as applied to the generalized coordinates and momenta of a given dynamical system; this is known as the Generalized Uncertainty Principle representation (GUP) \cite{Maggiore1,Maggiore2,Maggiore3,Kempf1,Kempf2}, and when applied to the minisuperspace it reproduces a dynamics similar to that of Brane Cosmology (BC) \cite{BC}, a cosmological sector of ST.

Furthermore, both LQC and BC (the latter in the form of a Randall-Sundrum model \cite{RS1,RS2}) predict modified Friedmann equations for the evolution of the isotropic Universe, see in particular \cite{effLQC1,effLQC2} and \cite{RSrev} respectively. In models with zero curvature and no cosmological constants, these two modified Friedmann equations differ only for a sign, which however has a very deep implication: while in LQC a Big Bounce appears, connecting the expanding and the collapsing branches of the Universe, in BC the singularity survives and such two branches remain separated. 

The main focus of this paper is the comparison of the implementation of the two phenomenological approaches of PQM and GUP on a Bianchi I Universe (a spatially flat and fully anisotropic model), both on a semiclassical and on a quantum level. The leading idea of our analysis is that both approaches can be stated in terms of modified Heisenberg algebras for the position and momentum operators, with the important difference of a sign that seems to affect deeply the cosmological dynamics since it is reflected on the Friedmann equation for the isotropic Universe \cite{Battistideformed}. More specifically, we observe that the GUP, when applied to the minisuperspace, reproduces the BC dynamics only if it is formulated with a square root term similarly to PQM.

We show here that such an isomorphism of the two approaches traced above (apart from the peculiarity of the sign) is only a formal feature. In fact, we demonstrate that PQM is never associated with a minimal uncertainty on the position operator, which instead is the basic morphology of a GUP proposal of the kind presented in \cite{Kempf2}. Clearly, on a semiclassical level, the isomorphism can be pursued since it results simply in a different sign in the modified Poisson brackets; however, we see here that, regarding the presence of the singularity, the different sign affects the modified dynamics of a Bianchi I model in the same way as for the isotropic Universe. Note that we implement the modified algebra on the volume variable only, using instead the standard Poisson brackets for the anisotropic degrees of freedom. We also show that the correct identifications between Polymer Cosmology (PC) and LQC on one hand and between the GUP and BC on the other remain valid also if we use in the latter case the momentum representation in terms of a hyperbolic sine (as a parallelism with the polymer framework where a trigonometrical one naturally appears).

On a quantum level, we are able to rewrite the Wheeler-DeWitt equation in an isomorphic form in both the PQM and the GUP formulations by suitably choosing a substitution for the momentum variable and, in the GUP case, the correct measure for the scalar product. The analysis confirms the existence of a Big Bounce in the PQM description and of a singular cosmology in the GUP formulation also for the quantum wavepackets. The analysis of the standard deviation shows that the localized packets inevitably spread, but this fact has different implications in the two frameworks: while for the polymer analysis the initial conditions can be set so that the wavepacket is peaked at the Bounce and then we can have a quasi-classical description across it, in the GUP case the existence of a minimal uncertainty on position (in our setting the volume variable) forces the treatment of the singularity as a full quantum phenomenon.

The paper is organized as follows. In Sec. \ref{altquantum} we present the different representations that we are going to use: we first review PQM and the GUP in their original setups, and then we introduce alternative formulations that make them comparable. In Sec. \ref{semicl} we present the classical dynamics of the Bianchi I model, and then implement the approaches described before to obtain the semiclassical evolution for all cases. Then in Sec. \ref{quantumcosm} we implement on Bianchi I on a quantum level both PQM and the GUP, but only in their original representations; we first show the inevitable spreading of the wavepackets and then study the expectation values of the volume operator in the two cases. In Sec. \ref{concl} we give a summary and some concluding remarks.

We work in the natural units $\hslash=c=8\pi G=1$.

\section{Alternative Representations of Quantum Mechanics}
\label{altquantum}
In this section we present two alternative representations of quantum mechanics, PQM and the GUP, and their variants. The main aim is to find the modified uncertainty principles and their consequences, as well as a viable formalism for the implementation of these representations to cosmology.

\subsection{Polymer Quantum Mechanics}
\label{POLY}
The aim of PQM is to implement a fundamental scale in the Hilbert space through the introduction of a lattice structure, and when applied to cosmology it yields a non-zero minimum for the volume of the Universe and replaces the Big Bang singularity with a Big Bounce  \cite{Mantero,Review}. As already mentioned, it is often considered as some kind of a low-energy limit of LQC (and indeed the two cosmological theories present many similarities, see for example \cite{Federico}), but it is derived independently and allows for more freedom in the choice of the variable to discretize. We will present the polymer representation following that in \cite{CorichiPQM}.

\subsubsection{Kinematics}
The starting point is a Hilbert space with abstract orthonormal kets $\ket{\mu_i}$ with $\braket{\mu_i}{\mu_j}=\delta_{ij}$, where $\mu_i$, $i=1,...,N$ are real numeric labels. In such a space it is possible to define two fundamental operators:
\begin{subequations}
\begin{equation}
\hat\epsilon\,\ket\mu=\mu\,\ket\mu,
\end{equation}
\begin{equation}
\hat s(\zeta)\,\ket\mu=\ket{\mu+\zeta},
\end{equation}
\label{polyabstractoperators}
\end{subequations}
respectively label and shift operator. The shift operator $\hat s(\zeta)$ is actually a family of parameter-dependent unitary operators, that due to orthonormality are discontinuous in $\zeta$ and cannot be generated by the exponentiation of a self-adjoint operator.

Let us now consider a Hamiltonian system with canonical variables $(q,p)$. In the momentum polarization, the fundamental states $\ket{\mu}$ become
\begin{equation}
\psi_\mu(p)=\braket{p}{\mu}=e^{i\mu p}.
\end{equation}
It is possible to prove that the Hilbert space of PQM in this polarization is given by $\mathcal{H}_\text{poly}=L^2(\mathbb{R}_B,d\mu_H)$, where $\mathbb{R}_B$ is the Bohr compactification of the real line and $d\mu_H$ is the Haar measure; note that this coincides with the kinematical Hilbert space of LQC \cite{LQC}. The two fundamental operators \eqref{polyabstractoperators} are identified respectively with the differential coordinate operator $\hat q$ and with the multiplicative operator $\hat T(\zeta)$:
\begin{subequations}
\begin{equation}
\hat q\,\psi_\mu(p)=-i\pdv{p}e^{i\mu p}=\mu\,\psi_\mu(p),
\end{equation}
\begin{equation}
\hat T(\zeta)\,\psi_\mu(p)=e^{i\zeta p}\,e^{i\mu p}=\psi_{\mu+\zeta}(p).
\end{equation}
\end{subequations}
$\hat T(\zeta)$ is now a discontinuous operator in $\mathcal{H}_\text{poly}$, and it is not possible to promote $p$ to a well-defined momentum operator; in order to implement a Hamiltonian constraint and derive the dynamics of a system, the momentum must be regularized.

\subsubsection{Regularization and Dynamics}
The regularization procedure consists in implementing a lattice with constant spacing on position, i.e. a regular graph $\gamma_{\mu_0}=\{q\in\mathbb{R}:q=\mu_n=n\mu_0\,\,\text{with}\,\,n\in\mathbb{Z}\}$ where $\mu_0\in\mathbb{R}$ is the fundamental polymer scale, and restricting the Hilbert space only to those states defined on the lattice:
\begin{equation}
\ket\psi=\sum_n\,b_n\,\ket{\mu_n},
\end{equation}
with $\sum_n\,\abs{b_n}^2<\infty$. Of course also the translational operator must be constrained to remain on $\gamma_{\mu_0}$ by setting $\zeta=\mu_0$:
\begin{equation}
\hat T(\mu_0)\,\ket{\mu_n}=\ket{\mu_{n+1}}.
\end{equation}
When the condition $\mu_0p\ll1$ is satisfied, we can write:
\begin{equation}
\label{ppoly}
p\approx\frac{\sin(\mu_0p)}{\mu_0}=\frac{e^{i\mu_0p}-e^{-i\mu_0p}}{2i\mu_0}
\end{equation}
and as a consequence we can define a regulated momentum operator through $\hat T(\mu_0)$:
\begin{equation}
\hat p_{\mu_0}\,\ket{\mu_n}=\frac{\hat{T}(\mu_0)-\hat{T}(-\mu_0)}{2i\mu_0}\,\ket{\mu_n}=\frac{\ket{\mu_{n+1}}-\ket{\mu_{n-1}}}{2i\mu_0}.
\label{regulatedp}
\end{equation}
Now it is possible to construct a Hamiltonian operator on the graph as ${\hat{\mathcal{C}}_{\gamma_{\mu_0}}=\frac{1}{2m}\hat{p}^2_{\mu_0}+\hat U(\hat q)}$, where $\hat{p}^2_{\mu_0}$ is the composition of the regularized operator \eqref{regulatedp} with itself (corresponding to a square sine approximation, although other definitions are possible \cite{CorichiPQM}) and $\hat U(\hat q)$ is the potential of the system under consideration.

When analyzing a system through the formalism of PQM, the differential coordinate operator and the regulated momentum operator \eqref{regulatedp} must be used. Alternatively, it is possible to derive a semiclassical evolution by using the formal substitution \eqref{ppoly} to obtain a polymer-modified Hamiltonian, thus including quantum modifications in the classical dynamics \cite{Mantero,Review,Antonini,MoriconiAniso_2017,MoriconiBianchiI_2018}. This is the usual way to implement PQM on a system; however, it is not the only possibility, as we will see in later section.

\subsection{Generalized Uncertainty Principle Representation}
The term GUP can be used to refer to any quantum theory that considers a modification of the standard Heisenberg Uncertainty Principle. In a one-dimensional system with variables $(q,p)$, the most analyzed uncertainty principle is expressed in general as \cite{Kempf1}
\begin{equation}
\Delta q\Delta p\geq\frac{1}{2}\,\Big(1+A(\Delta q)^2+B(\Delta p)^2\Big).
\end{equation}
We will restrict our analysis to the case $A=0$, that is of more interest to cosmology, and use the representation developed in \cite{Kempf2}. Note that the opposite case with $B=0$ and $A\neq0$ is sometimes referred to as Extended Uncertainty Principle or EUP, to distinguish between the two cases.

\subsubsection{Absolute Minimal Uncertainty on Position}
We consider the algebra generated by the modified commutation relations
\begin{equation}
\comm{q}{p}_\text{GUP}=i(1+Bp^2)\qq{with}B>0;
\label{kempfcommutator}
\end{equation}
through the general relation $\Delta O_1\Delta O_2\geq\frac{\abs{\ev{\comm{O_1}{O_2}}}}{2}$ valid for the symmetric operators $O_1$, $O_2$, it is easy to derive a GUP of the form
\begin{equation}
(\Delta q\Delta p)_\text{GUP}\geq\frac{1}{2}\Bigl(1+B\ev{p^2}\Bigr).
\label{kempfGUP}
\end{equation}
This expression implies a non-zero minimal uncertainty on position: using ${(\Delta p)^2=\ev{p^2}-\ev{p}^2}$ and solving for $\Delta p$, we obtain
\begin{equation}
\Delta p=\frac{1}{B}\left(\Delta q\pm\sqrt{(\Delta q)^2-B\Big(1+B\ev{p}^2\Big)\,}\right),
\end{equation}
and in order for the square root to be real we must have
\begin{equation}
\Delta q\geq\sqrt{B\Big(1+B\ev{p}^2\Big)\,}\,.
\end{equation}
Therefore, the absolute minimum uncertainty is obtained in the case $\ev{p}=0$:
\begin{equation}
\Delta q_0=\sqrt{B\,}\,.
\label{deltaq0GUP}
\end{equation}

An absolute minimal uncertainty implies that eigenstates of the position operator cannot exist: an eigenstate has by definition zero uncertainty (actually it is possible to define formal position eigenstates, but they have diverging energy and are not physical). Therefore a position polarization is not available, and we are forced to use the momentum polarization. On the other hand, the case $A\neq0$ implies a minimal uncertainty in momentum and prevents the use of that polarization; when both terms are present, a Bergmann-Fock construction must be used instead \cite{Kempf1}.

In the momentum representation states are wave functions $\psi(p)=\braket{p}{\psi}$, on which the fundamental operators of position and momentum act as
\begin{subequations}
\begin{equation}
\hat{p}\,\psi(p)=p\,\psi(p),
\end{equation}
\begin{equation}
\hat{q}\,\psi(p)=i(1+Bp^2)\dv{p}\psi(p);
\end{equation}
\label{GUPoperatorsKempf}
\end{subequations}
as shown in \cite{Kempf2}, they are symmetric with respect to the GUP-modified scalar product
\begin{equation}
\braket{\psi_1}{\psi_2}_\text{GUP}=\int_{-\infty}^{+\infty}\frac{dp}{1+Bp^2}\,\psi_1^*(p)\psi_2(p).
\label{GUPscalarproduct}
\end{equation}
When implementing the GUP on the quantum cosmological minisuperspace, since we will be dealing with wavepackets that are solutions to Klein-Gordon wave equations, we will use a Klein-Gordon-like scalar product with the measure $dp\,(1+Bp^2)^{-1}$, so that the operators are still symmetric.

\subsubsection{Quasi-Position Wavefunctions}
As mentioned previously, in the GUP representation of quantum mechanics eigenstates of the position operator are not physical and a position polarization is not available. In order to recover information on position, we can define ``states of maximal localization (ml)" around a position $z$, i.e, states $\ket{\psi^\text{ml}}$ that satisfy
\begin{subequations}
\begin{equation}
\ev{\hat{q}}{\psi^\text{ml}}=z,
\end{equation}
\begin{equation}
(\Delta q)^2\big|_{\ket{\psi^\text{ml}}}=\ev{\Bigl(q^2-\ev{q}^2\Bigr)}{\psi^\text{ml}}=(\Delta q_0)^2,
\end{equation}
\end{subequations}
where the scalar products are those presented in eq. \eqref{GUPscalarproduct}. Using in the GUP the values that yield the minimal uncertainty $\Delta q_0$, i.e. $\ev{p}=0$ and $\Delta p=B^{-1/2}$, it is possible to find the expression of a normalized maximally localized state in momentum space:
\begin{equation}
\psi_z^\text{ml}(p)=\sqrt{\frac{2\sqrt{B\,}}{\pi(1+Bp^2)}\,}\,\,e^{-i\,\frac{z\arctan(\sqrt{B\,}\,p)}{\sqrt{B\,}}}.
\end{equation}
These states have finite energy and are therefore physical; however they are not orthonomal:
\begin{equation}
\braket{\psi_{z_1}^\text{ml}}{\psi_{z_2}^\text{ml}}=\frac{\sin(\pi y)}{\pi y\left(1-y^2\right)}\qq{with}y=\frac{z_1-z_2}{2\sqrt{B\,}}.
\end{equation}
If we now project a generic state $\psi(p)$ onto a maximally localized state, we obtain the so-called ``quasi-position (qp)" wavefunction:
\begin{equation}
\psi^\text{qp}(z)=\braket{\psi(p)}{\psi_z^\text{ml}(p)}=\sqrt{\frac{2\sqrt{B\,}}{\pi}\,}\int_{-\infty}^{+\infty}\frac{dp}{(1+Bp^2)^{\frac{3}{2}}}\,\,e^{i\frac{z\arctan(\sqrt{B\,}\,p)}{\sqrt{B\,}}}\,\psi(p).
\label{quasiposition}
\end{equation}
This is somewhat a generalization of a Fourier transform and is invertible. The state $\psi^\text{qp}(z)$ represents the probability amplitude for a particle to be maximally localized around a position $z$.

The GUP algebra admits a representation on quasi-position space: the actions of the two operators $\hat{q}$ and $\hat{p}$ can be defined as
\begin{subequations}
\begin{equation}
\hat{p}\,\psi^\text{qp}(z)=-\frac{\tan(i\sqrt{B\,}\,\pdv{z})}{\sqrt{B\,}}\,\psi^\text{qp}(z),
\end{equation}
\begin{equation}
\hat{q}\,\psi^\text{qp}(z)=\left(z-\sqrt{B\,}\,\tan\bigg(i\sqrt{B\,}\,\pdv{z}\bigg)\right)\psi^\text{qp}(z).
\end{equation}
\end{subequations}
For more details on the quasi-position representation and a deeper functional analysis, see \cite{Kempf2}.

\subsection{PQM as a Modified Algebra}
Another way to implement PQM on a Hamiltonian system is to use modified commutation relations inspired by \cite{Battistideformed} of the kind
\begin{equation}
\comm{q}{p}_\text{poly}=i\sqrt{1-\mu_0^2p^2\,}\,;
\label{commpolysqrt}
\end{equation}
a representation for $p$ that satisfies this algebra is the polymer representation \eqref{ppoly}. At a semiclassical level, this modified commutation relation reduces to modified Poisson brackets. So, when considering PQM as a modified algebra the Hamiltonian constraint is not modified through the polymer substitution \eqref{ppoly} as usual, but modified Poisson brackets are considered in order to derive the polymer semiclassical dynamics. Moreover, if we take into account the condition $\mu_0p\ll1$, the cosine can be expanded and the contributions of the higher terms of the resulting Taylor series can be neglected. For example, in \cite{Alberto} we analyzed the semiclassical dynamics of the flat Friedmann-Lemaitre-Robertson-Walker (FLRW) model resulting from the modified Poisson bracket $\pb{q}{p}\approx1-\frac{\mu_0^2p^2}{2}$ i.e. truncating the series expansion after the second term. This was done in order to perform a comparison with the GUP approach presented earlier. 

In this work we will perform similar analyses on the anisotropic Bianchi I model. In particular, we consider two polymer-modified commutation relations ("exact" and "truncated") and try to see what the resulting Polymer Uncertainty Principle (PUP) is and whether or not the implementation of a lattice implies a minimal uncertainty.

\subsubsection{Truncated Polymer Commutator}
The expanded commutator up to second order is
\begin{equation}
\comm{q}{p}_\text{PUPt}=i(1-\frac{\mu_0^2}{2}p^2),
\label{commPUPt}
\end{equation}
where the index ``t" stands for ``truncated". Performing the same analysis as the GUP, the resulting PUP for the truncated case reads
\begin{equation}
(\Delta q\Delta p)_\text{PUPt}=\frac{1}{2}\abs{\ev{\,\,\comm{q}{p}_\text{PUPt}}}=\frac{1}{2}\left(1-\frac{\mu_0^2}{2}\Bigl((\Delta p)^2+\ev{p}^2\Bigr)\right);
\end{equation}
solving again for $\Delta p$ and requiring the root to be real, we obtain
\begin{equation}
(\Delta q)^2\geq\frac{\mu_0^2}{2}\Bigl(\frac{\mu_0^2}{2}\ev{p}^2-1\Bigr).
\end{equation}
It is clear how the minimum value of the r.h.s., reached in the case $\ev{p}=0$, is negative and therefore the l.h.s. can reach zero: PQM does \emph{not} imply a minimal uncertainty on position.

The actions of the operators corresponding to the truncated commutation relations can be defined as
\begin{subequations}
\begin{equation}
\hat{p}\,\psi(p)=p\,\psi(p),
\end{equation}
\begin{equation}
\hat{q}\,\psi(p)=i\Bigl(1-\frac{\mu_0^2}{2}p^2\Bigr)\pdv{p}\,\psi(p).
\end{equation}
\end{subequations}

\subsubsection{Exact Polymer Commutator on Gaussian States}
We consider now the polymer-modified commutator
\begin{equation}
\comm{q}{p}_\text{PUPe}=i\cos(\mu_0p),
\label{commPUPe}
\end{equation}
where the index "e" means "exact" i.e. not Taylor-expanded. This modified algebra corresponds to \eqref{commpolysqrt} where we have written the operators in the polymer p-representation \eqref{ppoly}. Now we must find the expectation value of this commutator. Note that the expectation value of the cosine can be computed only if we assume a specific form for the states: here we will choose Gaussian states, i.e. normalized wavefunctions of the form $\psi(p)\propto e^{-\frac{p^2}{2\sigma^2}}$ such that the result can be expressed as function of $\sigma^2=(\Delta p)^2$. This is not a loss of generality, because later in the (quantum) cosmological implementation we will study the evolution of wavepackets with Gaussian weights that can be seen as Fourier-transforms of Gaussian states. So, we will consider states of the form
\begin{equation}
\psi(p)=\frac{e^{-\frac{p^2}{2\sigma^2}}}{\sqrt{2\pi\sigma^2\,}}.
\label{genericGaussian}
\end{equation}
The integral and the resulting PUP are
\begin{equation}
\ev{\cos(\mu_op)}{\psi}=\int_{-\infty}^\infty\frac{dp}{4\pi\sigma^2}\,e^{-\frac{p^2}{\sigma^2}}\cos(\mu_0p)=\frac{e^{-\frac{\mu_0^2\sigma^2}{4}}}{2\sigma\sqrt{\pi\,}},
\end{equation}
\begin{equation}
(\Delta q\Delta p)_\text{PUPe}=\frac{\mu_0\,e^{-\frac{\mu_0^2\sigma^2}{4}}}{2\pi^\frac{1}{4}\sigma^\frac{1}{2}\sqrt{1-e^{-\mu_0^2\sigma^2}\,}}.
\end{equation}
It is clear how $\Delta q$ as function of $\sigma$ goes to zero in the limit $\sigma\to\infty$; therefore we confirm again that the PUP and thus PQM in general do not implement an absolute minimum uncertainty on position.

We remark that in \eqref{commPUPe} we have used the p-representation used in PQM, i.e. 
\begin{equation}
\hat{q}\,\psi(p)(p)=i\dv{p}\psi(p),\quad\hat{p}\,\psi(p)=\frac{\sin(\mu_0p)}{\mu_0}\,\psi(p).
\label{polyrep}
\end{equation}
However, it is also possible to use
\begin{equation}
\hat{q}\,\psi(p)=i\cos(\mu_0p)\dv{p}\psi(p),\quad\hat{p}\,\psi=p\,\psi(p)\,,
\end{equation}
i.e. a kind of representation more similar to the modified-algebra approaches, see \cite{Kempf1}. When implementing PQM to cosmology we will use representation \eqref{polyrep}.

\subsection{Generalized Uncertainty Principle from Brane-like Cosmology}
We have seen how PQM can be implemented as a modified algebra, and in particular how it can be made similar to the GUP through the truncated polymer commutator \eqref{commPUPt}, with the most relevant difference being a sign. It is possible to do the opposite, i.e. to implement a Generalized Uncertainty Principle in a form similar to pure PQM: if we suppose that $1+Bp^2$ are the first terms of a Taylor expansion, we can assume that the full commutator is
\begin{equation}
\comm{q}{p}_\text{GUPb}=i\cosh(\sqrt{2B\,}\,p),
\label{commGUPb}
\end{equation}
where ``b" stands for ``brane" since the implementation of this formalism to the cosmological minisuperspace reproduces effects similar to BC, as we will see in the next sections.

Assuming again normalized Gaussian states of the form \eqref{genericGaussian}, we can find the expectation value of \eqref{commGUPb} and the resulting GUP:
\begin{equation}
\ev{\cosh(\sqrt{2B\,}\,p)}{\psi}=\int_{-\infty}^\infty\frac{dp}{2\pi\sigma^2}\,e^{-\frac{p^2}{\sigma^2}}\cosh(\sqrt{2B\,}\,p)=\frac{e^{\frac{B\sigma^2}{2}}}{2\sigma\sqrt{\pi\,}},
\end{equation}
\begin{equation}
(\Delta q\Delta p)_\text{GUPb}=\frac{\sqrt{2B\,}\,e^{\frac{B\sigma^2}{2}}}{2\pi^\frac{1}{4}\sigma^\frac{1}{2}\sqrt{e^{2B\sigma^2}-1\,}};
\end{equation}
it is clear how also in this case $\Delta q$ goes to zero in the limit $\sigma\to\infty$.

Similarly to the PUPe case, there are (at least) two different representations that are able to reproduce the commutator \eqref{commGUPb}, both in momentum polarization:
\begin{subequations}
\begin{align}
1)\quad\hat{q}\,\psi(p)=i\cosh(\sqrt{2B\,}\,p)\dv{p}\psi(p),\quad\hat{p}\,\psi(p)=p\,\psi(p);\\
2)\quad\hat{q}\,\psi(p)=i\dv{p}\psi(p),\quad\hat{p}\,\psi(p)=\frac{\sinh(\sqrt{2B\,}\,p)}{\sqrt{2B\,}}\psi(p).
\end{align}
\label{GUPsinhrepresentation}
\end{subequations}
Note that representation 2 is very similar to the polymer substitution \eqref{ppoly}, but with a hyperbolic sine instead of a trigonometric one.

\section{Bianchi I Semiclassical Dynamics}
\label{semicl}
In order to obtain a better understanding of the minisuperspace formulation of canonical quantum gravity in the metric approach, it is interesting to investigate the dynamics of homogeneous but anisotropic models, i.e. of the Bianchi Universes. Since the three spatial directions evolve independently, these models are described by three different independent scale factors $a_i(t)$, $i=1,2,3$ along the three spatial directions; however it is often more useful to describe the model in the so-called Misner variables $(\alpha,\beta_\pm)$ that make the kinetic term in the Hamiltonian diagonal, where $\alpha$ is linked logarithmically to the isotropic volume while $(\beta_+,\beta_-)$ parametrize the anisotropies. Here we will use a variation of the Misner variables, obtained by substituting $\alpha$ with a variable $v$ proportional to the actual volume; this choice is due to the fact that in PC it has been shown that in general the discretization of a logarithmic variable does not solve the singularity, while the discretization of the volume replaces it with a Big Bounce independent on initial conditions \cite{Mantero,Review,Antonini,Giovannetti,Crino}.

The general line element for the Bianchi models in the Misner-like variables $(v,\beta_\pm)$ reads as
\begin{equation}
ds^2=\text{N}^2dt^2-v^\frac{2}{3}\left(e^{2\beta}\right)_{ab}\omega^a\omega^b,
\end{equation}
where the matrix $\beta$ has the form $\beta_{ab}=\text{diag}(\beta_++\sqrt{3\,}\beta_-,\beta_+-\sqrt{3\,}\beta_-,-2\beta_+)$, N is the timeshift, and $\omega_a,\omega_b$ are the 1-forms specifying the isometry group under which the specific model under consideration is invariant. Thanks to homogeneity, all the quantities will depend only on time.

The action $\mathcal{S}$ and Hamiltonian $\mathcal{C}$ of the Bianchi Universes in vacuum are
\begin{subequations}
\begin{equation}
\mathcal{S}_\text{B}=\int dt\bigl(P_v\dot{v}+P_+\dot{\beta}_++P_-\dot{\beta}_--\text{N}\mathcal{C}_\text{B}\bigr),
\end{equation}
\begin{equation}
\mathcal{C}_\text{B}=\frac{-9P_v^2v^2+P_+^2+P_-^2}{\xi\,v}+\frac{v^\frac{4}{3}}{4}U_\text{B}(\beta_\pm),
\end{equation}
\end{subequations}
where $\xi$ is a numerical constant coming from the spatial integral in the Einstein Hilbert action, $(P_v,P_\pm)$ are the momenta conjugate to $(v,\beta_\pm)$, while $U_\text{B}(\beta_\pm)$ is the potential associated to the specific model and is equal to zero in the simplest case of Bianchi I.

\subsection{Classical Dynamics}
The classical Hamiltonian of the Bianchi I model filled with a free massless scalar field $\phi$ is simply
\begin{equation}
\mathcal{C}_\text{class}=\mathcal{C}_\text{classBI}+\mathcal{C}_\phi=\frac{1}{\xi}\left(-9P_v^2v+\frac{P_+^2+P_-^2}{v}\right)+\frac{\xi}{12}\rho_\phi v=0,\quad\rho_\phi=\frac{P_\phi^2}{2v^2}
\label{BIHclass}
\end{equation}
where $\rho_\phi$ is the energy density of the scalar field and $P_\phi$ its conjugate momentum; the scalar field has been rescaled (through the constant $\frac{\xi}{12}$) without loss of generality in order to obtain a more readable expression for the Friedmann equation. From the Hamiltonian constraint we immediately see that the momenta $P_\pm$ and $P_\phi$ are constants of motion; this will later allow for the scalar field to be able to play the role of an internal relational time, in order to have a better comparison between the semiclassical and quantum dynamics. The equations of motion for the volume and the anisotropies are
\begin{subequations}
\begin{equation}
\dot{v}=-\frac{18}{\xi}P_vv,
\label{vdotclass}
\end{equation}
\begin{equation}
\dot{\beta}_\pm=\frac{2}{\xi}\,\frac{P_\pm}{v};
\end{equation}
\end{subequations}
substituting the Hamiltonian constraint in eq. \eqref{vdotclass}, we obtain a Friedmann equation for the Bianchi I model of the form
\begin{equation}
H^2=\frac{1}{9}\frac{\dot{v}^2}{v^2}=\frac{\rho_\phi}{3}+\frac{4}{\xi^2}\,\frac{P_+^2+P_-^2}{v^2}=\frac{\rho_\phi+\rho_\text{a}}{3},\quad\rho_\text{a}=\frac{12}{\xi^2}\,\frac{P_+^2+P_-^2}{v^2}.
\label{Friedmannclass}
\end{equation}
The anisotropic momenta $P_\pm$ add a contribution to the total energy density that dictates the evolution. Indeed, differently from the isotropic FLRW model, the Bianchi models are dynamical even without any kind of matter-energy \cite{Primordial}.

At this point it is useful to consider the scalar field $\phi$ as internal time. In order to do this, we must reintroduce the constant N (until now implicitly set to 1) and fix the time gauge $\dot{\phi}=1$:
\begin{equation}
\dot{\phi}=\text{N}\pdv{\mathcal{C}_\text{BI}}{P_\phi}=\text{N}\frac{\xi}{12}\,\frac{P_\phi}{v}=1\quad\Rightarrow\quad\text{N}=\frac{12}{\xi}\,\frac{v}{P_\phi}=\frac{6}{\xi}\sqrt{\frac{2}{\rho_\phi}\,}.
\label{timegauge}
\end{equation}
The equations of motion in this new time gauge are
\begin{subequations}
\begin{equation}
\dv{v}{\phi}=-\frac{18}{\xi}P_vv\text{N},
\end{equation}
\begin{equation}
\dv{\beta_\pm}{\phi}=\frac{2}{\xi}\,\frac{P_\pm}{v}\text{N}=\sqrt{\frac{24}{\xi^2}\frac{\rho_{\text{a}_\pm}}{\rho_\phi}\,}=\beta_{1_\pm},\quad\Rightarrow\quad\beta_\pm(\phi)=\beta_{0_\pm}+\beta_{1_\pm}\phi,
\end{equation}
\end{subequations}
where we separated the two anisotropy densities as $\rho_{a_\pm}=\frac{12}{\xi^2}\frac{P_\pm^2}{v^2}$ and introduced the constants $\beta_{j_\pm}$ highlighting the linear evolution of the anisotropies with respect to the scalar field. Thanks to the Hamiltonian constraint the r.h.s. of the Friedmann equation in terms of $\phi$ becomes a constant, making its solution straightforward:
\begin{equation}
\left(\frac{1}{v}\dv{v}{\phi}\right)^2=\frac{216}{\xi^2}\left(1+\frac{\rho_\text{a}}{\rho_\phi}\right)=v_1^2,\quad\Rightarrow\quad v(\phi)=v_0e^{\pm v_1\phi},
\end{equation}
where $v_0$ comes from integration and $v_1$ is constant since both densities $\rho_\phi$ and $\rho_\text{a}$ scale as $v^{-2}$.

As shown later in Fig. \ref{vphiclassPQM}, the dynamics presents singularities: the plus sign represents an expanding universe with a Big Bang, while the minus sign implies a contraction towards a Big Crunch. We will now implement the alternative representations of quantum mechanics  on a semiclassical level.

\subsection{Semiclassical Polymer Cosmology}
\label{semiclassicalPC}
The first modification that we are going to use is the polymer substitution \eqref{ppoly}. Therefore the dynamics is derived through the standard Hamilton equations but from a modified Hamiltonian constraint:
\begin{equation}
\mathcal{C}_\text{PQM}=\mathcal{C}_\text{PQMBI}+\mathcal{C}_\phi=\frac{1}{\xi}\left(-9v\,\frac{\sin[2](\mu_0P_v)}{\mu_0^2}+\frac{P_+^2+P_-^2}{v}\right)+\frac{\xi}{12}\rho_\phi v=0.
\end{equation}
Since we are modifying only the volume term, the equations for the anisotropies will not be affected; in contrast, in \cite{Giovannetti} also the variables $\beta_\pm$ were discretized, and it was shown that in the Bianchi IX model the classical chaotic behaviour is removed for certain values of the ratio between the isotropic and anisotropic lattice parameters, but at the same time the volume variable was replaced by the logarithmic variable $\alpha$ that prevents the resolution of the singularity (see also \cite{Review} for a better comparison).

The quantum corrections are visible on the equation of motion for the volume and on the resulting Friedmann equation:
\begin{equation}
\dot{v}=-\frac{18}{\xi}\,\frac{\sin(\mu_0P_v)\cos(\mu_0P_v)}{\mu_0}\,v,
\end{equation}
\begin{equation}
(H_\text{PQM})^2=\frac{36}{\xi^2}\frac{\sin[2](\mu_0P_v)}{\mu_0^2}\bigl(1-\sin[2](\mu_0P_v)\bigr)=\frac{\rho_\phi+\rho_\text{a}}{3}\left(1-\frac{\rho_\phi+\rho_\text{a}}{\rho_\mu}\right),
\label{FriedmannPQM}
\end{equation}
where $\rho_\mu=\frac{108}{\mu_0^2\xi^2}$ is a critical density (coming from the polymer modifications) that adds a critical point in the evolution of the model; already here we can say that the singularity will be regularized, since the total energy density $\rho_\phi+\rho_\text{a}$ cannot diverge. Indeed, introducing the time gauge \eqref{timegauge}, the Friedmann equation rewrites in an easily solvable form:
\begin{equation}
\dv{v}{\phi}=-\frac{18}{\xi}\,\frac{\sin(\mu_0P_v)\cos(\mu_0P_v)}{\mu_0}\,v\text{N},
\end{equation}
\begin{equation}
\left(\frac{1}{v}\dv{v}{\phi}\right)^2=v_1^2\left(1-\frac{\rho_\phi+\rho_\text{a}}{\rho_\mu}\right)\quad\Rightarrow\quad v_\text{PQM}(\phi)=R_\mu\cosh(v_1\phi),
\end{equation}
where $R_\mu^2=\frac{\rho_\phi+\rho_\text{a}}{\rho_\mu}\,v^2$ is a constant.

\begin{figure}
    \centering
    \includegraphics[scale=0.7]{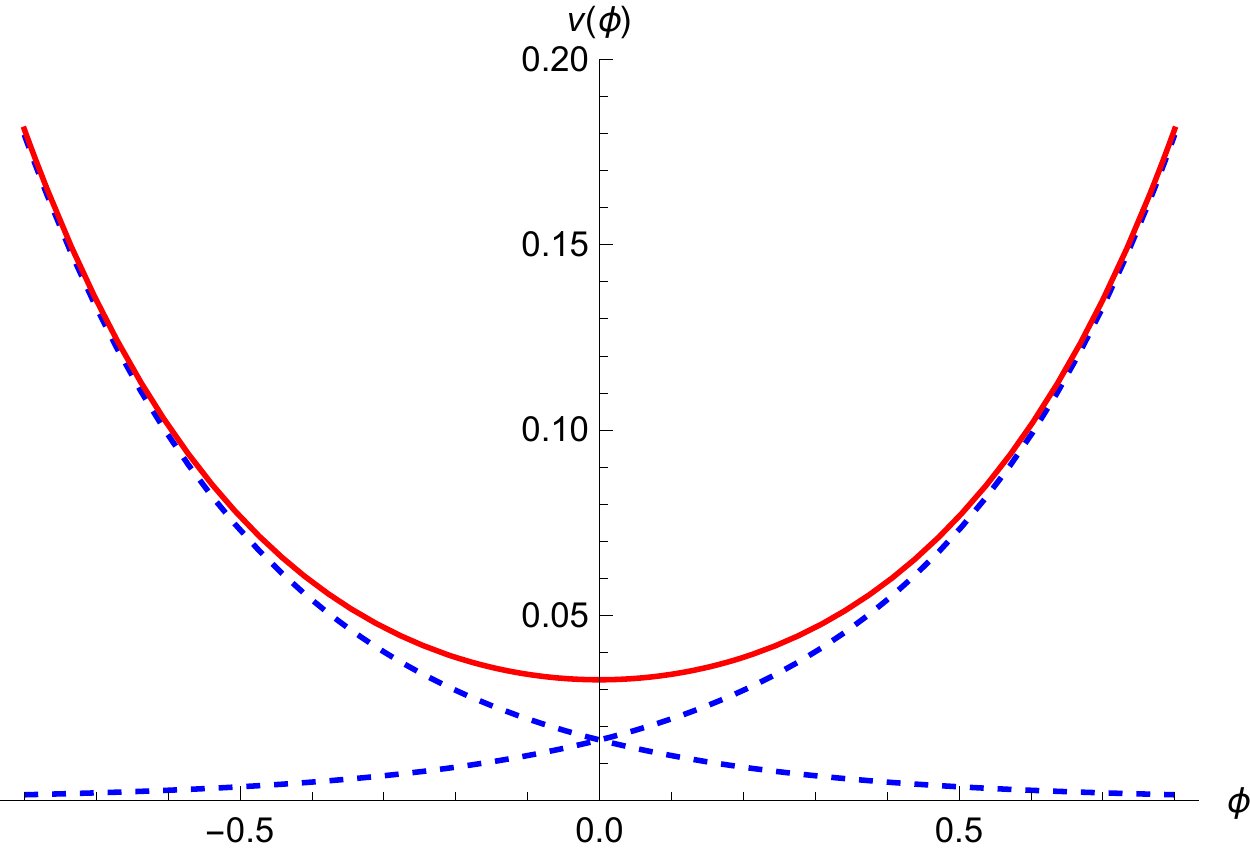}
    \caption{The evolution of $v(\phi)$ for the PQM case (red continuous line), compared with the classical case (blue dashed lines). The former jumps from the contracting classical branch to the expanding one, avoiding the singularity; at large times (both positive and negative) the classical dynamics is recovered.}
    \label{vphiclassPQM}
\end{figure}

As we can see from Fig. \ref{vphiclassPQM}, the PQM solution joins the two classical branches so that the singularity is replaced with a Big Bounce.

\subsection{Semiclassical GUP Dynamics}
\begin{figure}
    \centering
    \includegraphics[scale=0.7]{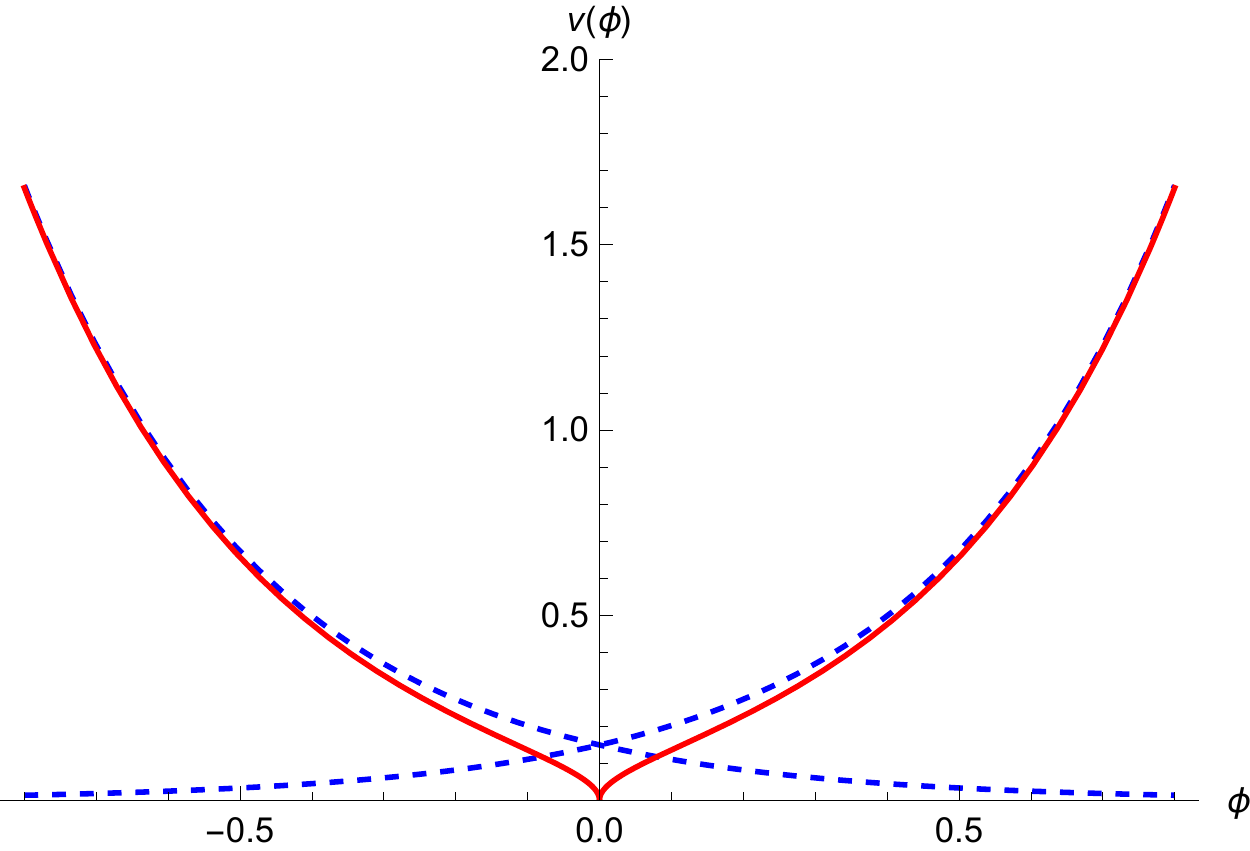}
    \caption{The evolution of $v(\phi)$ for the GUP case (red continuous line), compared with the classical case (blue dashed lines). The latter present a singularity at $\phi=0$; at large times (both positive and negative) the classical dynamics is recovered.}
    \label{vphiclassGUP}
\end{figure}

To implement GUP modifications at a semiclassical level, we will start from the unmodified Hamiltonian constraint \eqref{BIHclass} but derive the dynamics through modified Poisson brackets coming from the modified commutation relations \eqref{kempfcommutator}:
\begin{equation}
\mathcal{C}_\text{GUP}=\frac{1}{\xi}\left(-9P_v^2v+\frac{P_+^2+P_-^2}{v}\right)+\frac{\xi}{12}\rho_\phi v=0,\quad\pb{v}{P_v}_\text{GUP}=1+BP_v^2;
\end{equation}
also in this case the equations of motion for the anisotropies are not affected. The Hamilton equation for the volume yields a modified Friedmann equation:
\begin{equation}
\dot{v}=\pb{v}{P_v}\pdv{\mathcal{C}_\text{GUP}}{P_v}=-\frac{18}{\xi}P_vv(1+BP_v^2),
\end{equation}
\begin{equation}
(H_\text{GUP})^2=\frac{\rho_\phi+\rho_\text{a}}{3}\left(1+\frac{\rho_\phi+\rho_\text{a}}{2\rho_B}\right)^2,\quad\rho_B=\frac{54}{B\xi^2},
\end{equation}
where we defined the GUP critical density $\rho_B$ in this way in order to make the comparison with PQM and later approaches more evident. Here the correction factor, although similar to the one in eq. \eqref{FriedmannPQM}, has a different sign that will not introduce a critical point. Indeed, using the time gauge \eqref{timegauge}, we can find the evolution $v(\phi)$:
\begin{equation}
\dv{v}{\phi}=-\frac{18}{\xi}\,P_vv\text{N}(1+BP_v^2),
\end{equation}
\begin{equation}
\left(\frac{1}{v}\,\dv{v}{\phi}\right)^2=v_1^2\left(1+\frac{\rho_\phi+\rho_\text{a}}{2\rho_B}\right)^2,\quad v_\text{GUP}(\phi)=R_B\sqrt{e^{\pm2v_1\phi}-1\,},
\end{equation}
where $R_B^2=\frac{\rho_\phi+\rho_\text{a}}{2\rho_B}\,v^2=\,$const. We see that here there is no finite minimum for the volume because $v(\phi)$ does go to zero. As shown in Fig. \ref{vphiclassGUP}, the two solutions reach the singularity $v\to0$ as we expected.

\subsection{PUP-modified Dynamics}
As already mentioned, it is possible to implement polymer modifications through a modified algebra, leading to the two versions of the PUP. 

\begin{figure}
    \centering
    \includegraphics[scale=0.7]{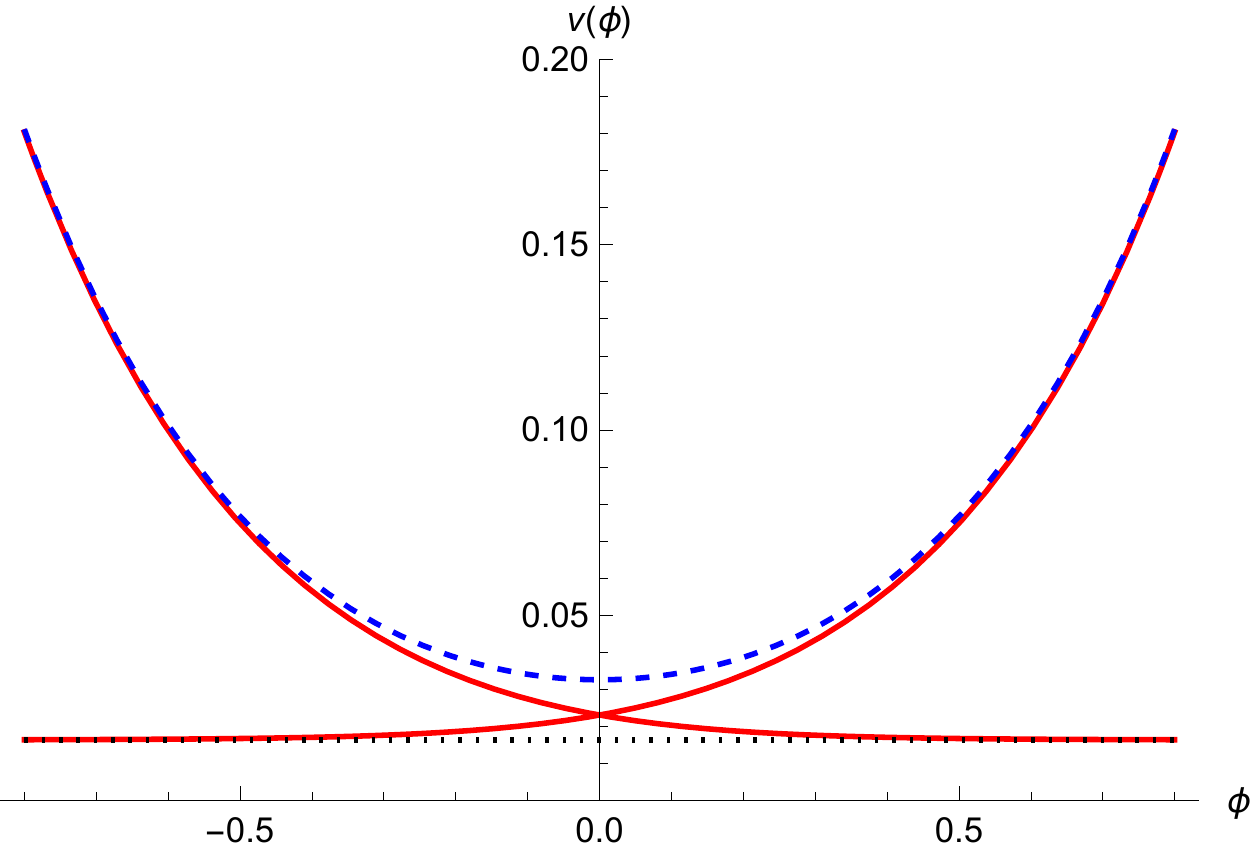}
    \caption{The evolution of $v(\phi)$ for the PUPt case (red continuous lines), compared with the full PQM case (blue dashed line). The former does avoid the singularity, but cannot reproduce the Bounce; at large times (both positive and negative) the PQM dynamics is recovered. The black dotted horizontal line highlights the asymptotic behaviour.}
    \label{vphiPUPtPQM}
\end{figure}

The two Poisson brackets that we are going to use to implement PUP modifications on the Bianchi I model are:
\begin{subequations}
\begin{equation}
\pb{v}{P_v}_\text{t}=\left(1-\frac{\mu_0^2}{2}P_v^2\right),
\label{pbPUPt}
\end{equation}
\begin{equation}
\pb{v}{P_v}_\text{e}=\sqrt{1-\mu_0^2P_v^2\,};
\label{pbPUPe}
\end{equation}
\end{subequations}
the starting point will be again the unmodified Hamiltonian constraint
\begin{equation}
\mathcal{C}_\text{PUP}=\mathcal{C}_\text{class}=\frac{1}{\xi}\left(-9P_v^2v+\frac{P_+^2+P_-^2}{v}\right)+\frac{\xi}{12}\rho_\phi v=0.
\end{equation}
The two equations of motion lead to two different modified Friedmann equations:
\begin{subequations}
\begin{equation}
(H_\text{t})^2=\frac{\rho_\phi+\rho_\text{a}}{3}\left(1-\frac{\rho_\phi+\rho_\text{a}}{2\rho_\mu}\right)^2,
\label{FriedmannPUPt}
\end{equation}
\begin{equation}
(H_\text{e})^2=\frac{\rho_\phi+\rho_\text{a}}{3}\left(1-\frac{\rho_\phi+\rho_\text{a}}{\rho_\mu}\right);
\end{equation}
\end{subequations}
as we can see, the exact case is identical to the PQM case, while the truncated case is slightly different but mimics it at low energies (if we expand the square, the first term is the same as the exact case). However, we must indeed remember that the truncated expansion is valid only in the low-energy regime $\mu_0P_v\ll1$; therefore the behaviour of $v_\text{PUPt}(\phi)$ may be different from that of $v_\text{PQM}(\phi)$ when we approach the singularity: solving the modified Friedmann equation \eqref{FriedmannPUPt} we have
\begin{equation}
v_\text{PUPt}(\phi)=R_\mu\sqrt{e^{\pm2v_1\phi}+1\,}.
\end{equation}
Indeed the truncation eliminates too many terms and is not able to reproduce the Bounce of full PQM as the square root does. This result corrects what was obtained in \cite{Alberto}. As we can see in Fig. \ref{vphiPUPtPQM}, the singularity is still avoided but, instead of with a Bounce, with an asymptote depending on the parameter $\mu_0$.

We therefore conclude that the two terms in the modified Poisson brackets \eqref{pbPUPt} are not sufficient to reproduce the Bounce of PQM. The addition of higher-order terms in the truncated Poisson bracket does not change the situation: each solution gets closer to the corresponding branch of the Bounce (i.e. the solution with the plus sign to the expanding branch, the solution with the minus to the collapsing one), but in the high-energy regime near the Bounce they are different and are never able to reproduce the full polymer dynamics.

On the other hand, the square root in \eqref{pbPUPe} reproduces the same dynamics of full PC; therefore, in later sections, we will study only the standard quantum PQM formalism.

\subsection{Semiclassical GUPb Cosmology}
The last modification that we are going to implement on the cosmological minisuperspace comes from the GUPb. There are two ways to do this: the first is to construct a modified Hamiltonian through the formal substitution with the hyperbolic sine coming from the representation \eqref{GUPsinhrepresentation}, in a way similar to semiclassical PC in section \ref{semiclassicalPC}; the second is to use the classical Hamiltonian but rewrite the modified commutator \eqref{commGUPb} analogously to the PUPe Poisson brackets \eqref{pbPUPe}, obtaining
\begin{equation}
\comm{q}{p}_\text{GUPb}=i\sqrt{1+2Bp^2}.
\end{equation}
Indeed, in \cite{Maggiore3} it is argued that a GUP with a square root term is actually more fundamental than the simple quadratic term of eq. \eqref{kempfcommutator}, since this implicitly assumes spin-zero particles.

The two GUPb cases are as follows:
\begin{subequations}
\begin{align}
1)\quad&\mathcal{C}_{\text{GUPb}_1}=\frac{1}{\xi}\left(-9v\,\frac{\sinh[2](\sqrt{2B\,}\,P_v)}{2B}+\frac{P_+^2+P_-^2}{v}\right)+\frac{\xi}{12}\rho_\phi v=0,\\
&\pb{v}{P_v}_{\text{GUPb}_1}=1,\\
2)\quad&\mathcal{C}_{\text{GUPb}_2}=\frac{1}{\xi}\left(-9P_v^2v+\frac{P_+^2+P_-^2}{v}\right)+\frac{\xi}{12}\rho_\phi v=0;\\
&\pb{v}{P_v}_{\text{GUPb}_2}=\sqrt{1+2BP_v^2\,}.
\end{align}
\end{subequations}
The equations of motion for the two cases, through the use of the corresponding Hamiltonian constraints, yield the same modified Friedmann equation:
\begin{equation}
(H_1)^2=(H_2)^2=\frac{\rho_\phi+\rho_\text{a}}{3}\left(1+\frac{\rho_\phi+\rho_\text{a}}{\rho_B}\right).
\end{equation}
This is the same Friedmann equation of BC \cite{RSrev}, where the role of the regularizing density $\rho_B$ is played by the Brane tension (and of course we don't have other terms such as curvature or a cosmological constant).

In the usual time gauge \eqref{timegauge} it becomes easily solvable:
\begin{equation}
\left(\frac{1}{v}\dv{v}{\phi}\right)^2=v_1^2\left(1+\frac{\rho_\phi+\rho_\text{a}}{\rho_B}\right)\quad\Rightarrow\quad v_\text{GUPb}(\phi)=\pm R_B\sinh(v_1\phi).
\end{equation}
Fig. \ref{v(phi)GUPb,GUP} shows the comparison with the original GUP representation; the singularity is still not avoided and is reached at a finite value of time $\phi$.

\begin{figure}
    \centering
    \includegraphics[scale=0.7]{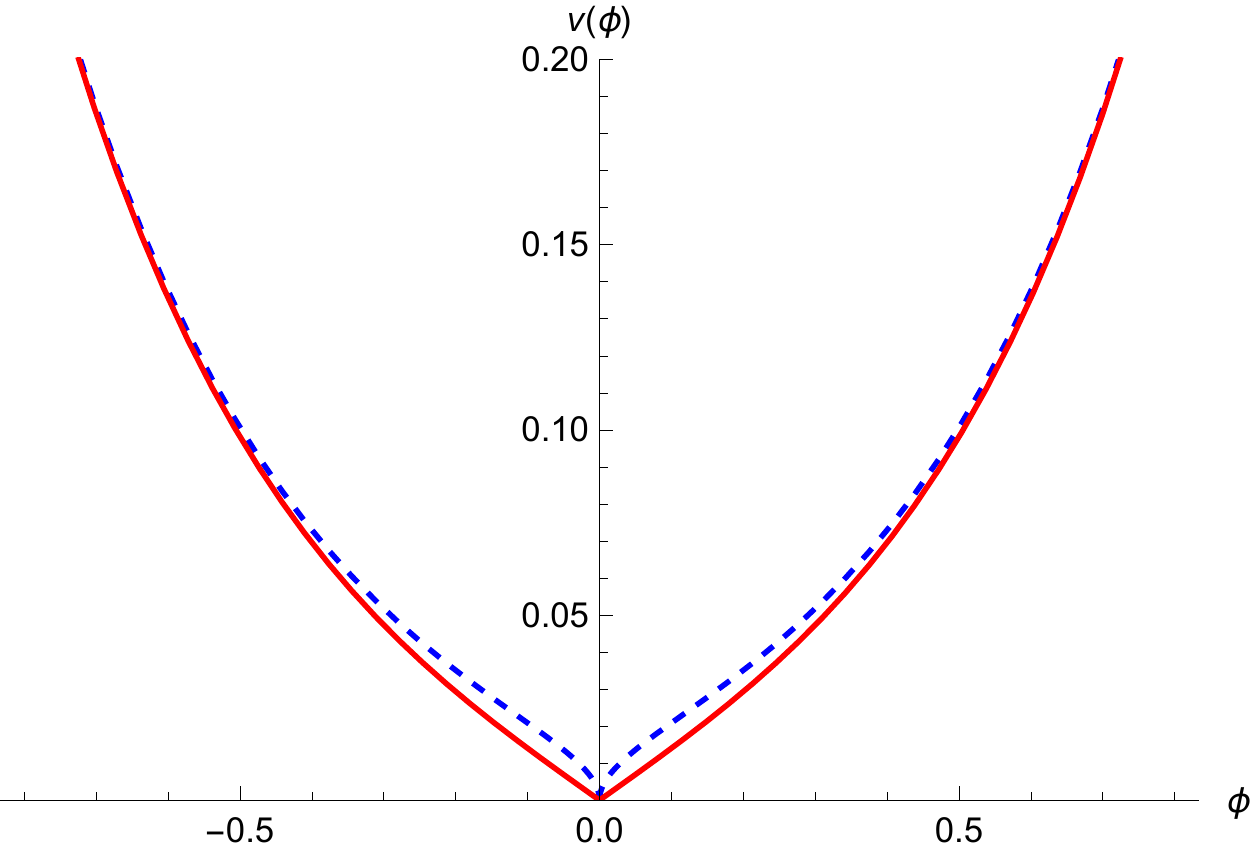}
    \caption{The evolution of $v(\phi)$ for the GUPb case (red continuous lines), compared with the original GUP case (blue dashed lines). The two are equivalent at large times (both positive and negative), and neither of them solves the singularity (although they approach it slightly differently).}
    \label{v(phi)GUPb,GUP}
\end{figure}

\section{Bianchi I Quantum Dynamics}
\label{quantumcosm}
In this section we will promote the fundamental variables to operators. Since we have shown that the PUPt formalism is not able to reproduce the full polymer dynamics while the PUPe formalism reproduces it exactly, and the GUPb modifications do not differ too much from the GUP ones, we will restrict the study to just full PQM and the original GUP. We will work in a hybrid polarization, i.e. momentum polarization for $v$ and position polarization (so to speak) for the other variables $\beta_\pm$ and $\phi$; this is due to the fact that in the two representations the position polarization is not viable: in PQM it would be non-trivial (see \cite{CorichiPQM}), while in the GUP it is not available and the quasi-position representation should be used instead.

The standard procedure is to perform some substitutions such that the WdW equations become Klein-Gordon-like wave equations where $\phi$ plays the role of time. The resulting solutions will be therefore expressed in a new variable $x_i$ with $i=$GUP, PQM and, although the substitutions will be different, the wavepackets as functions of $x$ will have the same form. When studying the physics, if we are interested in showing the spreading of the wavepacket, it will be the same in all pictures; however, the physical operators will of course have different expressions with respect to $x$ and their expectation values will be different, yielding inequivalent physical predictions in accordance with the semiclassical dynamics. This reflects the different nature of the coordinates in the two representations.

\subsection{Wavepacket Spreading}
\begin{figure}
\makebox[\textwidth]{
    \includegraphics[scale=0.75]{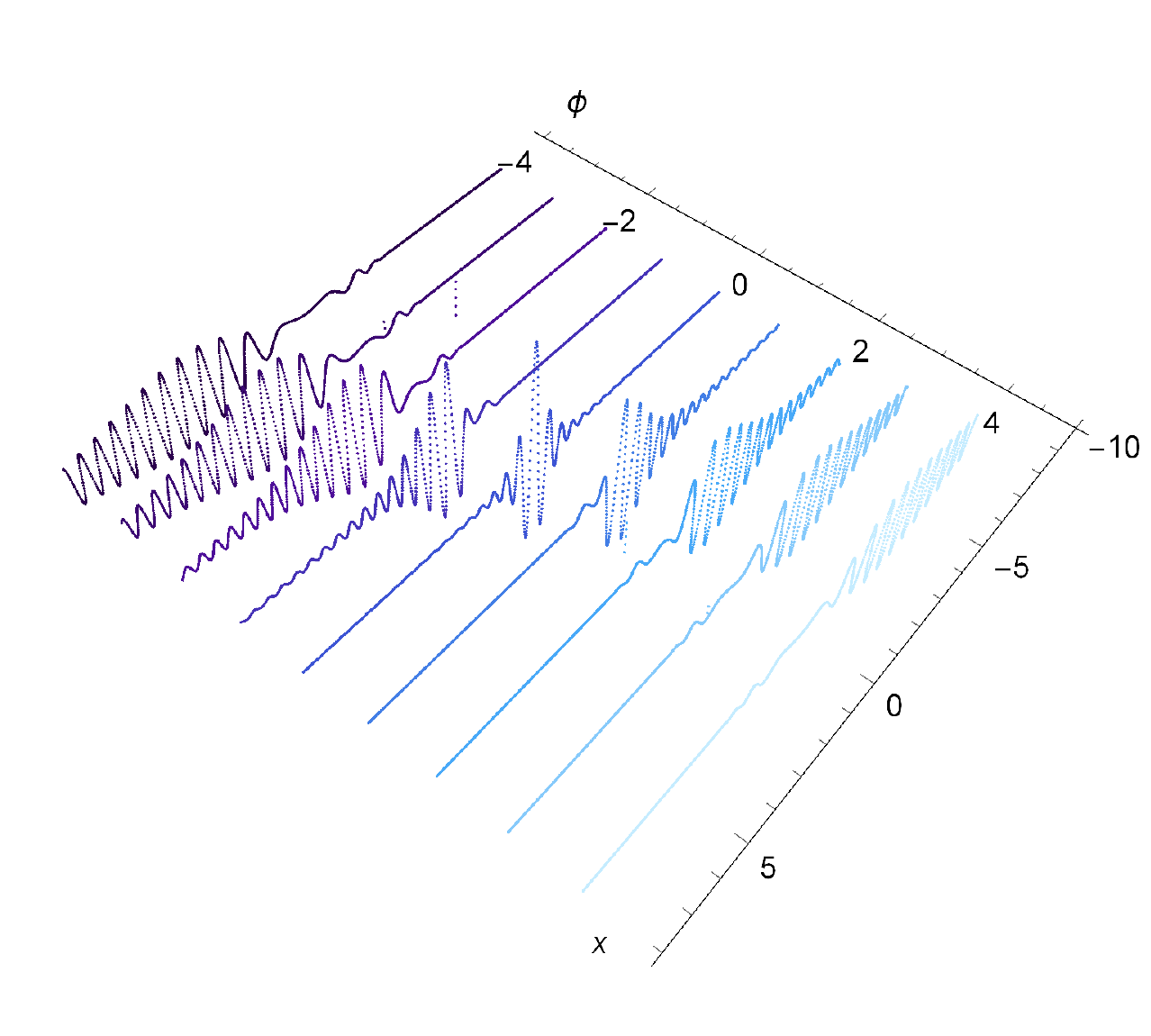}
    \includegraphics[scale=0.75]{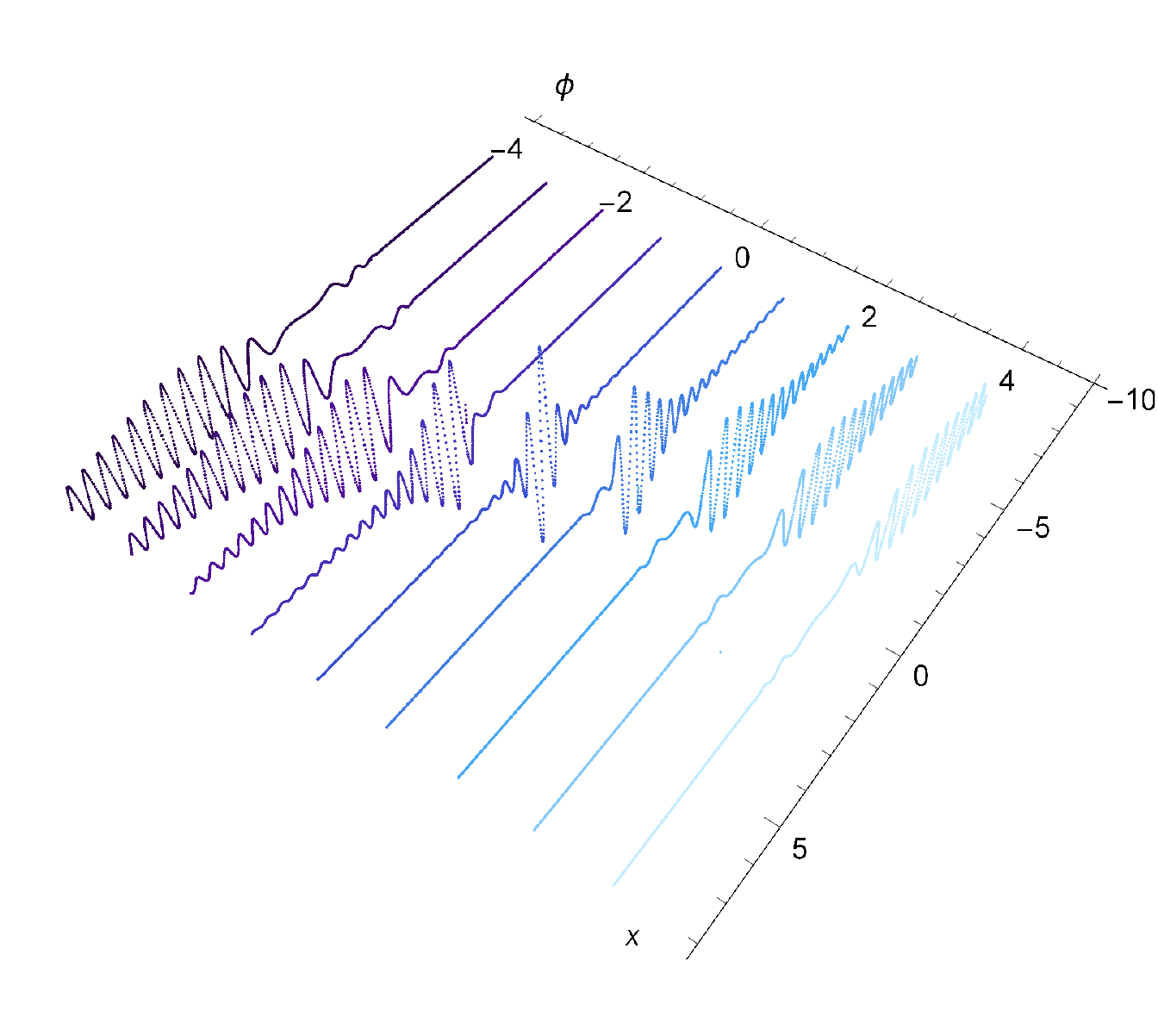}
}
\caption{The wavepacket $\Psi$ (real part on the left, imaginary part on the right) as function of the variable $x$ for different values of $\phi$. It is evident how it delocalizes at greater times.}
\label{pacchettospread}
\end{figure}

The Hamiltonian constraint, when promoted to a quantum level, yields a WdW equation that is the same in all cases:
\begin{equation}
(3\hat{P}_v\hat{v})^2\psi=(\hat{P}_+^2+\hat{P}_-^2+\hat{P}_\phi^2)\psi,
\label{WdWbruta}
\end{equation}
where we absorbed some constants in the scalar field momentum to lighten the notation (this is equivalent to a rescaling of the scalar field). The action of the operators not linked to the volume will always be the standard one (i.e. the variables $\beta_\pm$, $\phi$ will act multiplicatively and their conjugate momenta differentially); the differential volume operator and its conjugate momentum will have different expressions depending on the formalism used:
\begin{subequations}
\begin{align}
\text{PQM:}&\quad\hat{v}\,\psi=i\dv{P_v}\psi,\quad\hat{P_v}\,\psi=\frac{\sin(\mu_0P_v)}{\mu_0}\,\psi,\quad\comm{v}{P_v}=i\cos(\mu_0P_v);\\
\text{GUP:}&\quad\hat{v}\,\psi=i(1+BP_v^2)\dv{P_v}\psi,\quad\hat{P_v}\,\psi=P_v\psi,\quad\comm{v}{P_v}=i(1+BP_v^2);
\end{align}
\end{subequations}
it is easy to see that these definitions reproduce the corresponding modified commutators. The different actions will yield different terms on the l.h.s of eq. \eqref{WdWbruta}, and will therefore require different transformations for it to become a KG equation:
\begin{subequations}
\begin{align}
\text{PQM:}&\quad(3\hat{P}_v\hat{v})^2\psi=-\left(3\frac{\sin(\mu_0P_v)}{\mu_0}\,\dv{P_v}\right)^2\psi,\quad x_\text{PQM}=\frac{1}{3}\ln\abs{\tan(\frac{\mu_0P_v}{2})};\\
\text{GUP:}&\quad(3\hat{P}_v\hat{v})^2\psi=-\left(3P_v(1+BP_v^2)\dv{v}\right)^2\psi,\quad x_\text{GUP}=\frac{1}{3}\ln\abs{\frac{P_v}{\sqrt{1+BP_v^2\,}}};
\end{align}
\end{subequations}
\begin{equation}
\dv[2]{x_i}\psi=\left(\dv[2]{\beta_+}+\dv[2]{\beta_-}+\dv[2]{\phi}\right)\psi,\quad i=\text{ PQM, GUP}.
\label{KGx}
\end{equation}
Thanks to these substitutions, the WdW equation \eqref{WdWbruta} is rewritten in the KG form \eqref{KGx} and its general solution can be expressed as a wavepacket $\Psi$ with Gaussian weights $W$ peaked around specific values $\overline{k_\pm}$, $\overline{k_\phi}$ of the momenta eigenvalues that obey a non-linear dispersion relation:
\begin{subequations}
\label{wavepacket}
\begin{equation}
\Psi(x_i,\beta_\pm,\phi)=\int_{-\infty}^{+\infty}dk_+\,dk_-\,dk_\phi\,W(k_+)W(k_-)W(k_\phi)\psi(x_i,\beta_\pm,\phi,k_\pm,k_\phi),
\end{equation}
\begin{equation}
\psi(x_i,\beta_\pm,\phi,k_\pm,k_\phi)=e^{i(k_+\beta_++k_-\beta_-+k_\phi\phi+k_{x_i}x_i)}\,,
\end{equation}
\begin{equation}
W(k)=\frac{1}{\sqrt{2\pi\sigma^2\,}}\,e^{-\frac{(k-\overline{k})^2}{2\sigma^2}}\,,
\end{equation}
\begin{equation}
k_{x_i}=\sqrt{k_+^2+k_-^2+k_\phi^2\,\,}.
\end{equation}
\end{subequations}
Fig. \ref{pacchettospread} highlights the spreading of the wavepacket $\Psi$ with the passing of time $\phi$, as expected for a wavepacket propagating in more than one ``spatial" dimension (indeed, in a flat isotropic model where we only have $v$ and $\phi$ the wavepackets do not spread, as shown in \cite{Federico}). Note that this phenomenon restricts the validity of the quasi-classical limit of the quantum theory: if the wavepackets spread and delocalize, the quantum expectation values will follow the classical trajectory only for the brief interval when the packet is still localized enough.

\subsection{Quantum Expectation Values}
In this section, given the wavepacket \eqref{wavepacket}, we will compute the expectation value of the volume operator $\hat{V}=\hat{v}$ in the two different cases. In order to do this, we will use for PQM a KG scalar product of the form
\begin{equation}
\langle\hat{V}\rangle=\int_{-\infty}^{\infty}dx_id\beta_+d\beta_-\,\,i\left(\Psi^*\pdv{\phi}(\hat{V}\Psi)-(\hat{V}\Psi)\pdv{\phi}\Psi^*\right),
\end{equation}
while for the GUP representation the same measure as equation \eqref{GUPscalarproduct} is required for $\hat{V}$ to be symmetric.

\begin{figure}
\makebox[\textwidth]{
    \includegraphics[scale=0.7]{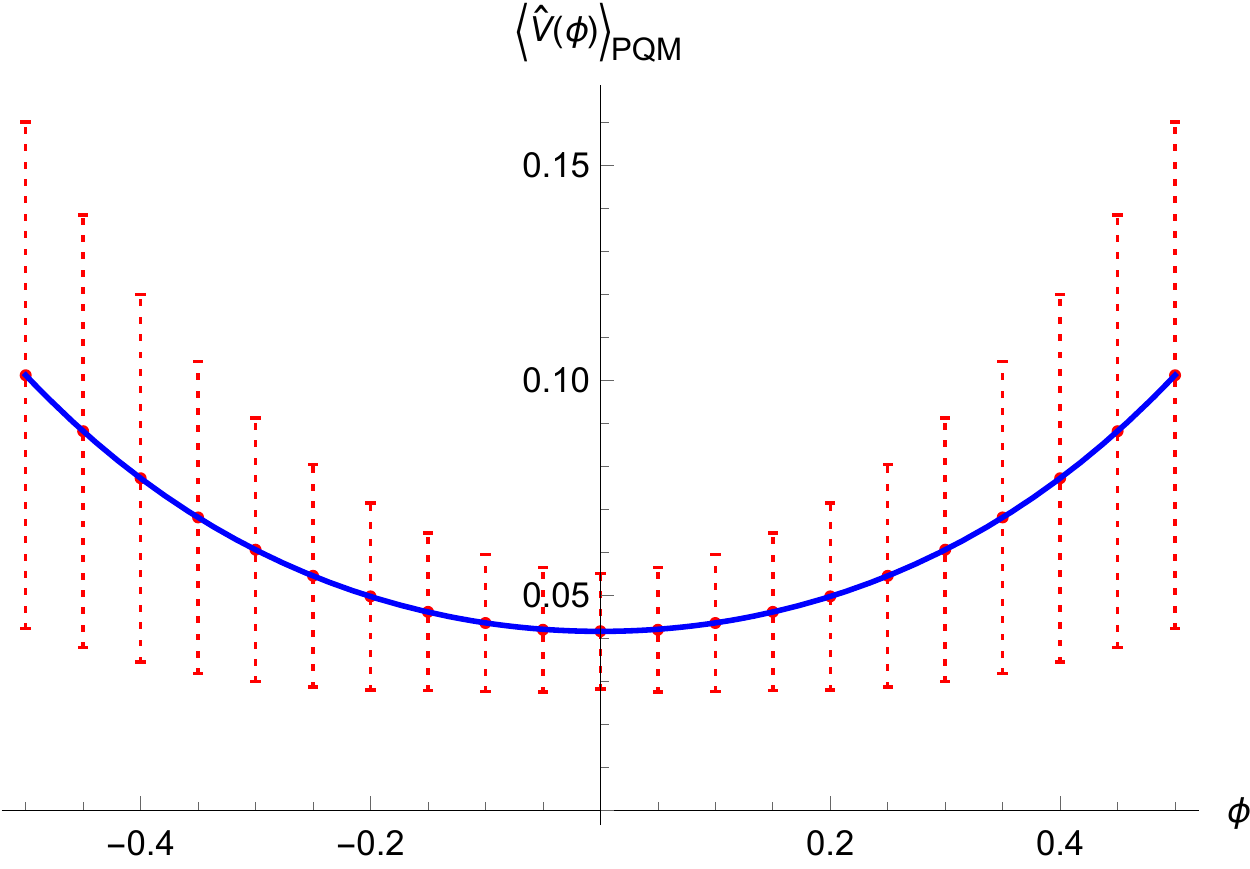}
    \quad
    \includegraphics[scale=0.7]{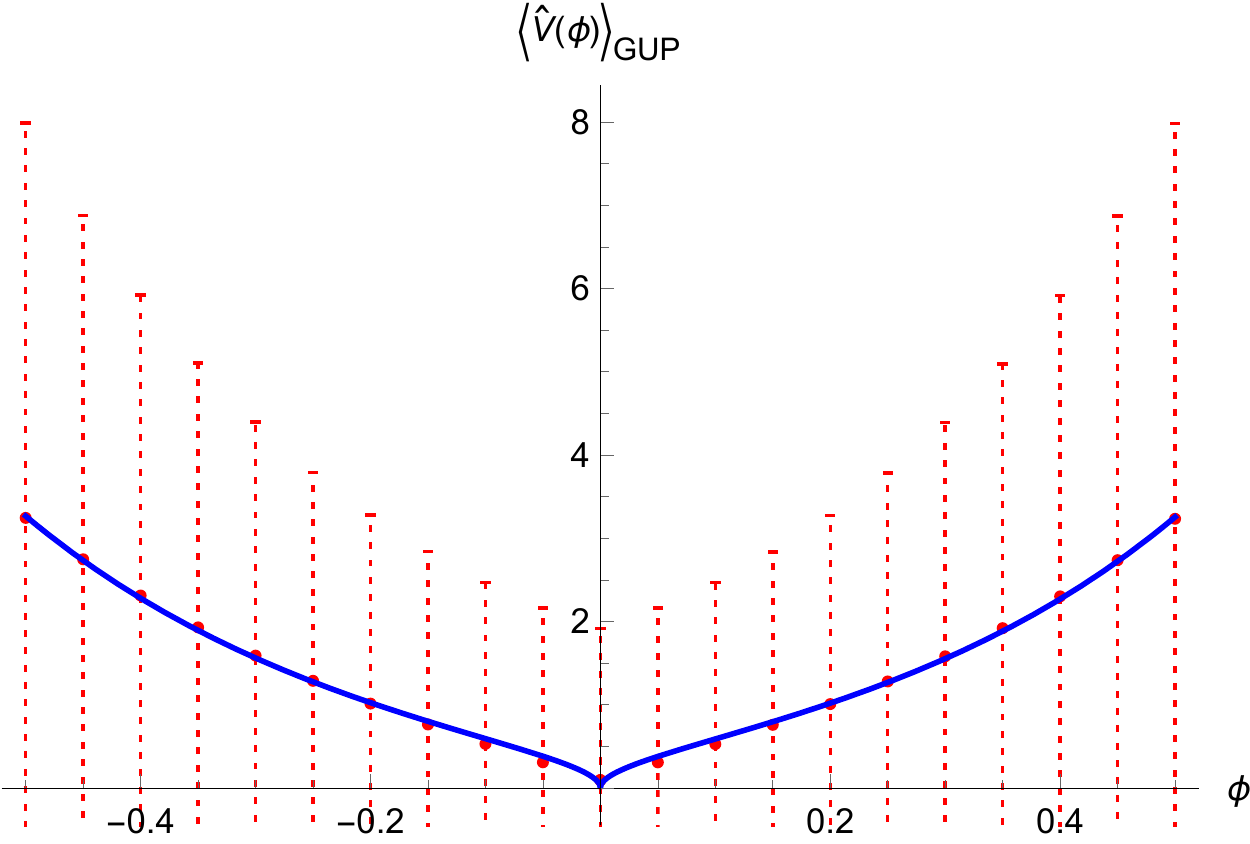}
}
\caption{Expectation value $\langle\hat{V}(\phi)\rangle$ of the volume as function of time for the quantum PQM formalism (left) and the quantum GUP formalism (right). The red dots have been computed numerically and fitted with the blue lines in accordance with the semiclassical trajectories; for the error bars we used the standard deviation $\sqrt{\langle\hat{V}^2\rangle-\langle\hat{V}\rangle^2}$.}
\label{QuantumMeanV}
\end{figure}

The volume operator will have a different form depending on the case:
\begin{subequations}
\begin{align}
\text{PQM:}&\quad\hat{V}=i\dv{P_v}=-\frac{i\mu_0}{3}\cosh(x)\dv{x},\\
\text{GUP:}&\quad\hat{V}=i(1+BP_v^2)\dv{P_v}=-\frac{i}{3}\frac{(1-Be^{6x})^\frac{1}{2}}{e^{3x}}\dv{x},
\end{align}
\end{subequations}
where we omitted the subscripts on $x$ to lighten notation. Note that the measure in the GUP case makes the power of the factor $1-Be^{6x}$ actually $3/2$ in the integral. In Figg. \ref{QuantumMeanV} we show the expectation value of the volume operator for the two different formalisms: the semiclassical trajectories are reproduced closely enough at small times, but the standard deviations start to grow quite soon; therefore these expectation values are reliable only in a small interval of time.

\section{Concluding remarks}
\label{concl}
We presented a detailed comparison of the semiclassical and quantum implementations of PQM and GUP to the evolution of a Bianchi I cosmology, outlining how this two approaches reproduce some basic features of LQG and ST respectively. In particular, the polymer approach reflects the concept of space graphs at the ground of LQG and we demonstrated that it is not associated to a minimal uncertainty on the position operator, differently from what the original formulation of the GUP predicts coherently with the string nature of a particle.

Then, on a semiclassical level we considered modified Poisson algebras implemented only on the Universe volume and we analyzed the morphology of the modified Friedmann equations of the Bianchi I model. The interesting result that emerged from this analysis is that the parallelism between the polymer approach versus LQC and the GUP one versus BC is guaranteed only if also the GUP-modified Poisson brackets are formulated with a square root, as first suggested in \cite{Maggiore1}; this strengthens the results obtained in \cite{Maggiore3}, where it is implied that a GUP with a square root term is more general and fundamental than the one introduced in \cite{Kempf1,Kempf2}. The present analysis generalizes that one in \cite{Battistideformed}, where a similar argument was developed for the isotropic Universe. We also showed that the same features emerge by recasting the GUP as somewhat a modified gravity theory, i.e. replacing the momentum variable in the classical Hamiltonian with a hyperbolic sine, similarly to what is usually done in PQM with a trigonometric sine. This paradigm strengthened the parallelism with LQC and BC, offering a new point of view on how the GUP procedure could be generalized to better describe a cosmological Brane theory of the minisuperspace.

Finally, we show how on a quantum level both the construction of the polymer lattice and the impossibility to deal with a vanishing uncertainty on the position operator implement cut-off effects in the early Universe dynamics, but they lead to a very different phenomenology. We have seen how the semiclassical polymer formulation is able to remove the singularity, replacing it with a Big Bounce, while the GUP formalism does not affect the singular nature of the Bianchi I model; this same feature is also confirmed by the pure quantum analysis, performed by implementing the two algorithms of cut-off physics according to their standard formulations \cite{CorichiPQM,Kempf2}. We constructed for both procedures the same Wheeler-DeWitt equation, although with a variable having very different meaning in the two cases; then, the evolution of Gaussian wavepackets has been investigated, showing that the expectation values as functions of the internal clock (i.e. the massless scalar field) do follow the classical trajectories, but sooner or later the states spread and delocalize. This implies that a pure quantum description of the Planckian regime is more meaningful in the case of an anisotropic model, especially in the GUP formulation where a minimal uncertainty on position prevents the localization of the wavepackets at will; on the other hand, in PQM this possibility is always available and the dynamics can present a quasi-classical morphology. As a consequence, the inability of the GUP representation to remove the singularity in the semiclassical description, contrary to PQM, is not conclusive, since a predictive quantum picture of the primordial Universe is still lacking. 

To conclude, the main goal of our analysis was to better characterize the physical meaning of PQM and GUP, especially in view of their cosmological implementation. The obtained equivalence of the effective Bianchi I dynamics with the corresponding LQC and BC predictions constitutes a significant starting point to extend the present picture and infer general features about the removal of the singularity in the so-called generic cosmological solution. Actually, the direct implementation of the two fundamental theories above would be forbidden by the complexity of the Superspace model, while the PQM and GUP versions of the so-called BKL conjecture are viable \cite{Antonini,BKL82}. However, in the ST the quantized gravitational modes arise in a perturbative scheme and the relic notion of a classical background still survives, differently from LQG. Moreover, the original formulation of the GUP introduces cut-off physics effects on a non-perturbative level in the low energy limit of ST, so this feature could suggest that the cosmological GUP implementation cannot fulfill all the features of a Planckian Universe in BC.

\end{document}